\documentclass[pra,twocolumn,superscriptaddress,showpacs,amsmath,amssymb,longbibliography,aps,10pt]{revtex4-2}
\usepackage{graphicx, color}
\usepackage{dcolumn}
\usepackage{bm}
\usepackage{epstopdf}
\usepackage{physics}
\usepackage[dvipsnames]{xcolor}
\usepackage{subfigure}
\usepackage{pifont}

\definecolor{orange}{rgb}{1,0.5,0}
\definecolor{goodgreen}{rgb}{0.1,0.5,0}
\definecolor{goodred}{rgb}{0.7,0,0}
\usepackage[colorlinks,urlcolor=goodgreen,citecolor=blue,linkcolor=goodred]{hyperref}
\usepackage{verbatim}
\usepackage[normalem]{ulem}
\graphicspath{ {./images/} }

\usepackage{lineno}

\usepackage{makecell}
\newcommand\hlinethickness[1]{\noalign{\hrule height #1}}

\let\oldepsilon\epsilon \let\epsilon\varepsilon \let\varepsilon\oldepsilon
\let\oldphi\phi \let\phi\varphi \let\varphi\oldphi

\begin{document}
\title{Characterization of a quantum bus between two driven qubits}

\begin{abstract}
We investigate the use of driven qubits coupled to a harmonic oscillator to implement a $\sqrt{i\mathrm{SWAP}}$-gate. By dressing the qubits through an external driving field, the qubits and the harmonic oscillator can be selectively coupled, allowing for the measurement of individual qubit states, as well as leading to effective qubit-qubit interactions. We compare the qubit readout on bare and dressed qubits, and demonstrate that when coupled to low-frequency resonators, dressed qubits provide a more robust readout than bare qubits in the presence of damping and thermal effects. Furthermore, we study the impact of various system parameters on the fidelity of the two-qubit gate, identifying an optimal range for quantum computation. Our findings guide the implementation of high-fidelity quantum gates in experimental setups, for example those employing nanoscale mechanical resonators.
\end{abstract}

\author{Alberto Hijano}
\email{alhijano@jyu.fi}
\affiliation{Department of Physics and Nanoscience Center, University of Jyvaskyla, P.O. Box 35 (YFL), FI-40014 University of Jyväskylä, Finland}

\author{Henri Lyyra}
\affiliation{Department of Physics and Nanoscience Center, University of Jyvaskyla, P.O. Box 35 (YFL), FI-40014 University of Jyväskylä, Finland}
\affiliation{Department of Mechanical and Materials Engineering, University of Turku, Turun yliopisto FI-20014, Finland}

\author{Juha T. Muhonen}
\affiliation{Department of Physics and Nanoscience Center, University of Jyvaskyla, P.O. Box 35 (YFL), FI-40014 University of Jyväskylä, Finland}

\author{Tero T. Heikkil\"a}
\email{tero.t.heikkila@jyu.fi}
\affiliation{Department of Physics and Nanoscience Center, University of Jyvaskyla, P.O. Box 35 (YFL), FI-40014 University of Jyväskylä, Finland}

\maketitle

\section{Introduction}
\label{sec:introduction}

Universal quantum computation, i.e., the ability to construct general quantum circuits, is a major milestone in quantum technologies. Some of its key challenges include scaling stable qubits, minimizing noise, implementing error correction protocols, increasing qubit coherence times, and reducing quantum gate operation times~\cite{Almudever:2017,Corcoles:2020}. A set of universal quantum gates enables the execution of any unitary operation using only the gates from that set. Having a small number of distinct gate types in a universal quantum computer is desirable, as it simplifies design, error correction, and control mechanisms required for reliable quantum computation. Two-qubit gates have been proven to be universal~\cite{DiVincenzo:1995}; for example, a CNOT gate or a $\sqrt{i\mathrm{SWAP}}$-gate, together with single-qubit rotation gates, form a universal set of quantum gates~\cite{quantum-computing-book,Barenco:1995}. Another key aspect of quantum computing is the ability to measure the final state of qubits accurately. Both high-fidelity gates and readout mechanisms are essential for practical quantum computing.

Several works have theoretically and experimentally demonstrated the selective coupling of arbitrary pairs of qubits via a common data bus~\cite{You:2001,Makhlin:2001,You:2002,Migliore:2003,Blais:2003,Plastina:2003,Sillanpaa:2007,Majer:2007}. Typically, this coupling mechanism requires the characteristic frequencies of the qubits and the bus to be of the same order. However, in many physical implementations, the characteristic frequencies of the qubits~\cite{Muhonen:2015,Laucht:2017} and the harmonic oscillator~\cite{Pigeau:2015,Muhonen:2019} may differ by orders of magnitude. One approach to coupling qubits with transition frequencies that differ from the bus frequency is to apply an external drive, dressing the qubit and effectively shifting the dressed qubit's transition frequency into resonance with the bus~\cite{Liu:2006,Eastham:2013,Laucht:2017}. This method offers two advantages: first, it provides a mechanism to control the qubit-qubit interaction time required to construct a desired quantum gate; second, it allows for the selective coupling of arbitrary qubit pairs.

\begin{figure}[t!]
    \centering
    \includegraphics[width=0.99\columnwidth]{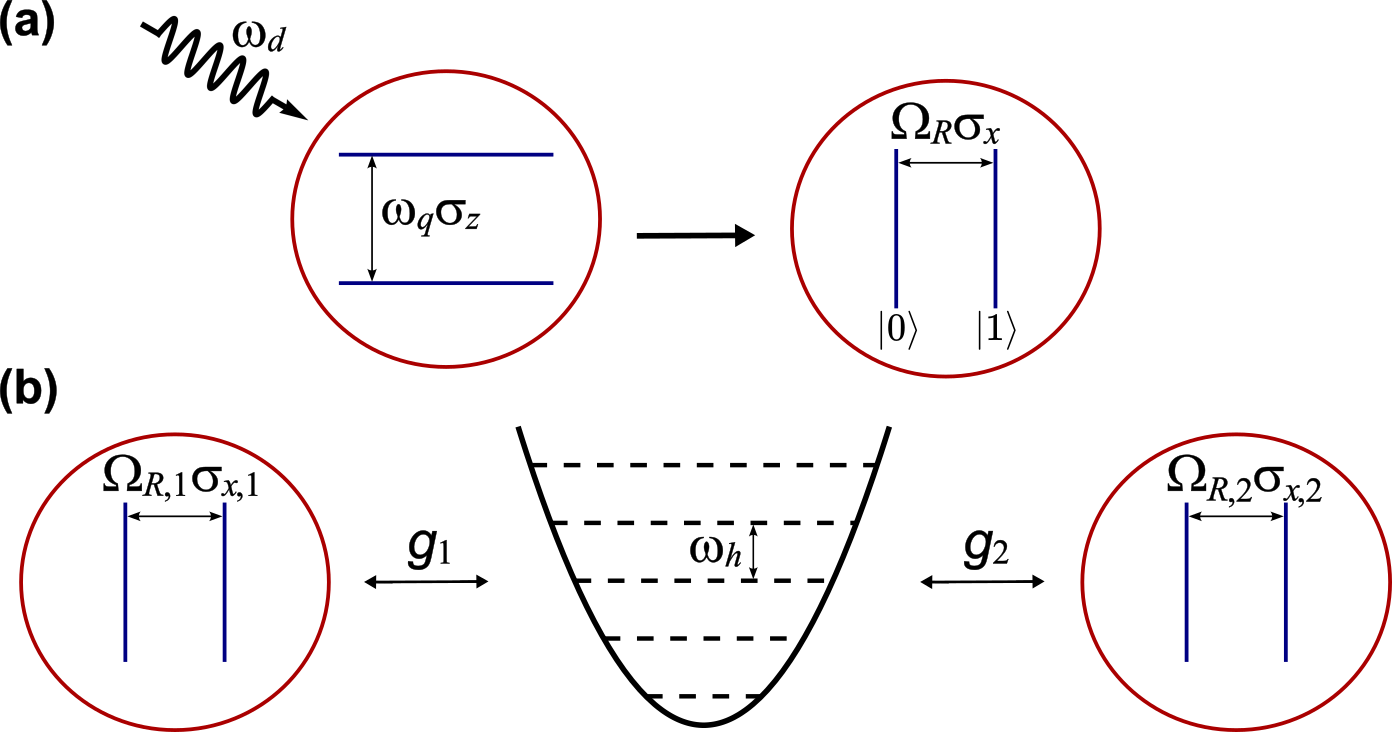}
    \caption{(a) Dressing of a qubit through a driving field with frequency $\omega_d$ and amplitude $\Omega_R$. (b) Sketch of the harmonic-oscillator mediated qubit-qubit coupling.}
    \label{fig:sketch}
\end{figure}

In this work, we compare the effects of dissipation and thermal effects on bare and dressed qubits coupled to a harmonic oscillator. By dressed qubit, we refer to a qubit driven by an AC field, with the computational basis defined in the frame rotating with the drive frequency, as described in Sec.~\ref{sec:dressed}. First, we analyze how the qubit's state can be inferred by measuring a resonance frequency shift of the harmonic oscillator. In the dispersive regime, such a measurement corresponds to a quantum nondemolition measurement of the qubit. This mechanism has been demonstrated in transmission line resonators~\cite{Blais:2004,Schuster:2005,Wallraff:2005,Gambetta:2006,Gambetta:2008}. The effect of dissipation on the system due to its coupling to the environment, modeled as a heat bath, depends on the dressing of the qubit. Interestingly, we find that the qualitative form of dissipation differs from that of bare qubits: whereas in bare qubits the steady state is the ground state, corresponding to one of the qubit eigenstates, a driven qubit, whose eigenstates are given in the rotating frame, relaxes toward a mixed state. The latter makes the two logical qubit states symmetric in terms of the unwanted dissipation. This may be advantageous for practical data processing protocols, since asymmetric logical states require more complex, asymmetric error models. We also demonstrate that the state readout mechanism is more robust for a dressed qubit compared to a bare one, particularly at higher temperatures. Following the approach of Refs.~\cite{Blais:2004,Liu:2006,Blais:2007,Srinivasa:2024}, we study how the harmonic-oscillator-mediated coupling can be utilized to implement a two-qubit gate. Earlier studies lack a thorough analysis of how thermal effects and other non-idealities influence gate fidelity. We show how the thermal occupation of the harmonic oscillator affects the ability to generate qubit-qubit entanglement. Additionally, we compare gate fidelities across different model parameters and propose an optimal parameter range for practical implementation. Notably, we show that the fidelity exhibits a non-monotonic dependence on the dressing rise time. This non-monotonic behavior can be leveraged to enhance fidelity in setups with a finite dressing speed.

The coupling of qubits through a common quantum bus has been demonstrated in superconducting charge qubits coupled via a microwave transmission line~\cite{Sillanpaa:2007,Majer:2007}, operating in the GHz range. The rate at which qubits exchange information is governed by the coupling strength: a higher coupling constant results in faster operations, but can also lead to stronger unwanted interactions and increased sensitivity to noise. As a result, shielding from charge noise may impose a limit on the operational speed of electromagnetic buses. An alternative implementation of a quantum bus can be achieved by coupling dressed spin qubits to a mechanical resonator~\cite{Rosenfeld:2021,Rabl:2009,Laucht:2017}, which work on the MHz scale. The spins couple to the motion of the mechanical resonator if the qubit transition frequency is comparable to the oscillator's frequency. Dressing enables the selective coupling of qubits to the quantum bus, while the interaction strength between the spin and the mechanical resonator can be controlled either by the intrinsic strain of the resonator or by a magnetic field gradient~\cite{Lyyra:2024}. Using low frequency resonators makes it feasible to match the resonator and Rabi frequencies, since the Rabi frequency increases monotonically with the microwave power. On the negative side, using a low-frequency resonator can lead to a significant thermal population in the resonator. In Sec.~\ref{sec:realization}, we justify the choice of the parameters used in this work by comparing them to a practical implementation employing nanomechanical resonators.

The work is organized as follows. In Sec.~\ref{sec:model}, we introduce the quantum master equation formalism used to study the qubit-harmonic oscillator system. In Sec.~\ref{sec:characterization}, we investigate how the damping and temperature affect the qubit state readout mechanism, comparing the performance of dressed and bare qubits and highlighting the advantages of using dressed qubits. In Sec.~\ref{sec:gate}, we analyze how various system parameters influence the fidelity of a two-qubit gate mediated by a harmonic oscillator and propose an optimal parameter range for practical implementation. In Sec.~\ref{sec:realization}, we discuss possible realizations of the qubit-harmonic oscillator system across different technological platforms. Finally, we summarize the results in Sec.~\ref{sec:conclusions}.

\section{Model and formalism}
\label{sec:model}

In this section, we introduce the open quantum system formalism used to study the interaction between a dressed qubit and a harmonic oscillator. Quantum systems are inherently coupled to an uncontrollable environment, leading to undesirable effects in quantum technologies, such as damping and thermal excitation. To describe this coupling, we apply the quantum master equation (QME) approach. The QME is the most common method for studying open quantum systems; it effectively traces out the environmental degrees of freedom and replaces them with terms that describe pure dephasing and the excitation and relaxation of system modes.

\subsection{Dressed qubit-harmonic oscillator coupling}\label{sec:dressed}
We study a qubit with transition frequency $\omega_q$, coupled to a harmonic oscillator with resonance frequency $\omega_h$. The Hamiltonian describing the coupling is (here and below $\hbar=k_B=1$)
\begin{equation}\label{eq:bare qubit}
    \hat{H}_\mathrm{bare}=\frac{\omega_q}{2}\hat{\sigma}_z+\omega_h\hat{a}^\dagger \hat{a}+g\hat{\sigma}_x(\hat{a}+\hat{a}^\dagger)\; ,
\end{equation}
where $\hat{\sigma}_i$ are the Pauli matrices describing the qubit degrees of freedom, $\hat{a}^\dagger$ and $\hat{a}$ are the raising and lowering operators of the harmonic oscillator, respectively, and $g$ is the coupling strength, which depends on the coupling mechanism and its implementation. As described in Appendix~\ref{sec:spring effect}, within the rotating wave approximation, the coupling term describes the exchange of energy quanta between the qubit and the harmonic oscillator. If $\omega_q$ is comparable to $\omega_h$, the coupling between the subsystems results in a significant interaction. However, the characteristic frequencies of the qubit and the bus may differ by several orders of magnitude. If the detuning between the qubit and the harmonic oscillator is very high $\omega_q-\omega_h\gg g$, the qubit may be driven by an input field in resonance with $\omega_q$ to dress the qubit with the desired eigenfrequency of the Rabi oscillations $\Omega_R$, so that the dressed qubit is in near resonance with the oscillator $\Omega_R\sim\omega_h$. For instance, a spin qubit can be dressed by applying a microwave field. Its magnetic field component, pointing in a direction perpendicular to the dc field, dresses the spin qubit, as shown in Fig.~\ref{fig:sketch}(a). In this work we consider a low-frequency bus $\omega_h\ll\omega_q$.

Here, we treat the driving field as a classical field interacting with the qubit. The Hamiltonian describing the dressed qubit-harmonic oscillator system then reads:
\begin{equation}
    \hat{H}_\mathrm{dress}=\frac{\omega_q}{2}\hat{\sigma}_z+\Omega_R(t)\cos{(\omega_d t)}\hat{\sigma}_x+\omega_h\hat{a}^\dagger \hat{a}-g\hat{\sigma}_z(\hat{a}+\hat{a}^\dagger)\; ,
\end{equation}
where $\Omega_R$ and $\omega_d$ are the amplitude and frequency of the drive. Moreover, in order to accomplish an off-diagonal coupling between the dressed qubit and the harmonic oscillator and thereby allow for exchanging quanta between the two systems, we now assume that the oscillator couples to the bare qubit eigenstates diagonally (via the $\hat{\sigma}_z$ term instead of the $\hat{\sigma}_x$ term). In Sec.~\ref{sec:QME} we describe how to account for the damping and the thermal excitations introduced by the interaction.

If the driving field detuning $\Delta=\omega_d-\omega_q$ is small compared to $\omega_d$ and $\omega_q$, we may work in the frame rotating with the drive, described by the unitary transformation $\hat{U}=e^{i\omega_d t \hat{\sigma}_z/2}$, and use the rotating-wave approximation to disregard the doubly rotating terms. This leads to $\hat{\tilde{H}}_\mathrm{dress} = \hat U \hat H_\mathrm{dress} \hat U^\dagger+i\partial_t \hat U \hat U^\dagger$ with
\begin{equation}\label{eq:dressed Hamiltonian}
    \hat{\tilde{H}}_\mathrm{dress}=-\frac{\Delta}{2}\hat{\sigma}_z+\frac{\Omega_R(t)}{2}\hat{\sigma}_x+\omega_h\hat{a}^\dagger \hat{a}-g\hat{\sigma}_z(\hat{a}+\hat{a}^\dagger)\; ,
\end{equation}
where the first two terms in the Hamiltonian describe the dressed qubit. For a drive in resonance with the bare qubit $\Delta=0$, and within the rotating wave approximation (note that this is distinct from the rotating wave approximation mentioned for the qubit-harmonic oscillator coupling), Eq.~\eqref{eq:dressed Hamiltonian} becomes the well-known Jaynes–Cummings Hamiltonian, see Appendix~\ref{sec:spring effect}. In this work, the logical states of the dressed qubit $\{|0\rangle,|1\rangle\}$ are given by the eigenstates of the driven Hamiltonian in the rotating frame (eigenstates of $\hat{\sigma}_x$), as defined in Table~\ref{table_basis}. The qubit can be selectively coupled to the harmonic oscillator by changing the driving amplitude $\Omega_R(t)$. In Sec.~\ref{sec:fidelity}, we study how the qubit coupling speed affects the fidelity of a harmonic oscillator-mediated gate.

As shown in Appendix~\ref{sec:spring effect}, if the dressed qubit is near-resonant with the harmonic oscillator, $\Omega_R \sim \omega_h$, the coupling between them shifts the resonance frequency of the oscillator in a way that depends on the qubit state. Therefore, the state of the qubit can be detected by measuring the noise power spectral density of the $\hat{x}(t)=\hat{a}+\hat{a}^\dagger$ quadrature
\begin{equation}\label{eq:Sx}
    S_x(\omega)=\int dt e^{i\omega t}\frac{\langle\{\hat{x}(t),\hat{x}(0)\}\rangle}{2}\; .
\end{equation}
The resonance frequency manifests as a maximum in $S_x(\omega)$, while the damping rates of the qubit and the harmonic oscillator determine its linewidth.

\begin{table}[t!]
    \centering
    \begin{tabular}{|c!{\vrule width 1pt}c|c|c|}
     \hline
      & Bare qubit & \makecell{Dressed qubit\\(bare basis)} & \makecell{Dressed qubit\\(dressed basis)}\\
     \hlinethickness{1pt}
     \makecell{Transition\\ frequency} & $\omega_q$ & $\Omega_R$ & $\Omega_R$ \\ 
     \hline
     \makecell{Hamiltonian} & $\omega_q\hat{\sigma}_z$ & $\Omega_R\hat{\sigma}_x$ & $\Omega_R\hat{\sigma}'_z$ \\
     \hline
     \makecell{Resonance\\ condition} & $\omega_q\sim\omega_h$ & $\Omega_R\sim\omega_h$ & $\Omega_R\sim\omega_h$ \\
     \hline
     \makecell{Logical\\ states} & \makecell{$|1\rangle=|\!\uparrow\rangle$\\$|0\rangle=|\!\downarrow\rangle$} & \makecell{$|1\rangle=\frac{1}{\sqrt{2}}(|\!\uparrow\rangle+|\!\downarrow\rangle)$\\$|0\rangle=\frac{-1}{\sqrt{2}}(|\!\uparrow\rangle-|\!\downarrow\rangle)$} & \makecell{$|1\rangle=|\!\uparrow\rangle'$\\$|0\rangle=|\!\downarrow\rangle'$} \\ [1ex] 
     \hline
    \end{tabular}
    \caption{Comparison between the physical quantities describing the bare and dressed qubits. The eigenstates of the dressed qubit are given in the rotating frame.}
    \label{table_basis}
\end{table}

\subsection{Quantum master equation}
\label{sec:QME}
The Lindblad quantum master equation may be used to describe quantum systems coupled to a heat bath~\cite{Gardiner-Collett-input-output}. The QME formalism accounts for both thermal excitation and damping arising from the system's interaction with the heat bath. This approach is valid when the coupling between the system and the bath is weak and when the subsystems interacting with the bath each have a single characteristic frequency---a condition that holds for two-level systems and harmonic oscillators. In this section, we introduce the QME for a dressed qubit coupled to a harmonic oscillator. The total Hamiltonian is thus given by
\begin{equation}\label{eq:total Hamiltonian}
    \hat{H}=\hat{H}_\mathrm{sys}+\hat{H}_\mathrm{B}+\hat{H}_{\gamma}
\end{equation}
where $\hat{H}_\mathrm{sys}$ is the system Hamiltonian, given by Eq.~\eqref{eq:bare qubit} for a bare and Eq.~\eqref{eq:dressed Hamiltonian} for a dressed qubit. $\hat{H}_\mathrm{B}$ describes the bosonic heat baths coupled to the qubit and the harmonic oscillator, while $\hat{H}_{\gamma}$ represents a linear interaction between the system and the bath.

The evolution of the total density matrix (including the system and the baths) is described by the von Neumann equation:
\begin{equation}
    \dot{\hat{\rho}}_\mathrm{tot}=-i[\hat{H},\hat{\rho}_\mathrm{tot}]\; ,
\end{equation}
where the total Hamiltonian is given by Eq.~\eqref{eq:total Hamiltonian}. For a reservoir weakly coupled to the system, and assuming that the dynamics of the environment are much faster than that of the system, we take the Born-Markov approximation~\cite{Gardiner_book}. Tracing the bath degrees of freedom out, the density matrix for the system obeys the QME:
\begin{equation}\label{eq:QME}
    \dot{\hat{\rho}}=-i[\hat{H}_\mathrm{sys},\hat{\rho}]+\sum_k\left(\hat{L}_k\hat{\rho}\hat{L}_k^\dagger-\frac{1}{2}\{\hat{L}_k^\dagger\hat{L}_k,\hat{\rho}\}\right)\; .
\end{equation}
$\hat{L}_k$ are Lindblad operators describing the effect of the baths on the system. They are related to the system operators coupling to the baths as 
\begin{subequations}\label{eq:Lindblad}
\begin{align}
\hat{L}_{q,-}&=\sqrt{(n_{\mathrm{th},q}+1)\gamma_q}\hat{\sigma}_-\\
\hat{L}_{q,+}&=\sqrt{n_{\mathrm{th},q}\gamma_q}\hat{\sigma}_+\\
\hat{L}_{h,-}&=\sqrt{(n_{\mathrm{th},h}+1)\gamma_h}\hat{a}\\
\hat{L}_{h,+}&=\sqrt{n_{\mathrm{th},h}\gamma_h}\hat{a}^\dagger,
\end{align}
\end{subequations}
where $n_{\mathrm{th},q}=(e^{\omega_q/T}-1)^{-1}$ and $n_{\mathrm{th},h}=(e^{\omega_h/T}-1)^{-1}$ describe the thermal state of the bath fields coupling to the qubit and the harmonic oscillator, respectively. Here, we assume that the qubit-harmonic oscillator coupling and the drive are sufficiently weak, so that they do not alter the Lindblad operators. Note that the system-bath coupling $\hat{H}_{\gamma}$ and the Lindblad operators are modified in the rotating frame considered for the dressed qubit. For a linear bath coupling, the qubit's Lindblad operators acquire a time-dependent phase factor in the rotating frame $\hat{\tilde{L}}_{q,\pm}=e^{\pm i\omega_d t}\hat{L}_{q,\pm}$, which does not affect the Lindblad master equation~\eqref{eq:QME}.

For the dressed qubit, we assume that the bare transition frequency is much higher than the temperature, i.e., $\omega_q \gg T$, so that $n_{\mathrm{th},q}\approx0$. The terms containing the Lindblad operators describing the coupling of the harmonic oscillator to the heat bath, $\hat{L}_{h,\pm}$, induce damping, driving the harmonic oscillator toward a thermal equilibrium state with the heat bath. Similarly, the terms containing $\hat{L}_{q,\pm}$ thermalize a bare qubit with its environment. Since we are interested in analyzing thermal effects, we assume that the qubit damping is stronger than pure dephasing, and only consider the Lindblad operators $\hat{L}_{q,\pm}$ describing the damping. Neglecting pure dephasing is justified, for example, for some nuclear spins in NV centers~\cite{Suter:2025}.

In the low temperature limit $T\ll\omega_q$, the bare qubit evolves toward its ground state in the steady-state regime. However, for a dressed qubit, the operators $\hat{\sigma}_\pm$ no longer describe energy quanta exchange with the environment, and the system rather evolves toward a mixed state~\cite{atom-photon-interactions-book}. A dressed qubit ideally decays into a maximally mixed state, but as shown in Sec.~\ref{sec:characterization}, the coupling with a harmonic oscillator results in a steady state with a negative $\langle\hat{\sigma}_x\rangle$. The interaction between the qubit and the harmonic oscillator couples the $|1,n=0\rangle$ and $|0,n=1\rangle$ states, where the first index corresponds to the dressed qubit and the second index denotes the oscillator's quantum number. This coupling transfers population from the $|1,n=0\rangle$ state to the $|0,n=1\rangle$ state, enforcing the negative $\langle\hat{\sigma}_x\rangle$.

\begin{figure*}
    \centering
    \includegraphics[width=0.9\textwidth]{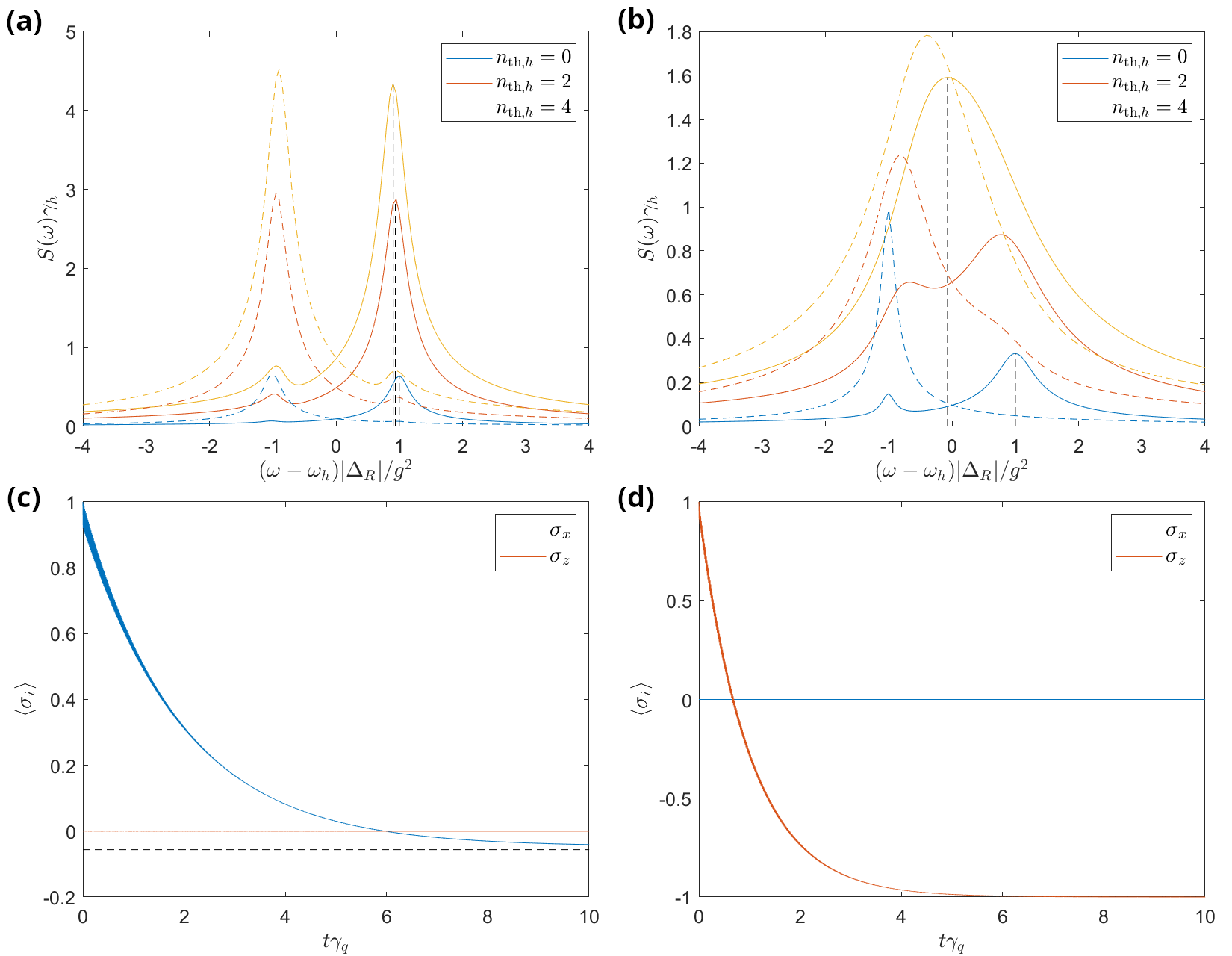}
    \caption{(a) Noise power spectral density for different temperatures for (a) a dressed qubit with $\omega_q=50\omega_h$, $\Delta=0$ and $\Delta_R=0.05\omega_h$, and (b) a bare qubit with $\omega_q=1.05\omega_h$. The solid lines correspond to the initial state $|1\rangle$ and the dashed lines to the initial state $|0\rangle$. The vertical dashed lines indicate the maximum of the spectral density. Time evolution of the qubit state for (c) dressed and (d) bare qubits, for thermal occupation $n_{\mathrm{th},h}=0$. All plots use the parameters $g=5 \cdot 10^{-3}\omega_h$, $\gamma_q=10^{-4}\omega_h$ and $\gamma_h=10^{-4}\omega_h$.}
    \label{fig:spring_effect}
\end{figure*}

\section{Characterization of the qubit readout mechanism}
\label{sec:characterization}
A quantum circuit consists of a sequence of quantum gates followed by measurements that yield information in the form of a classical bit (0 or 1), according to the Born rule. Qubit measurements are typically made in the qubit's eigenstate basis, which usually defines the computational basis $\{|0\rangle,|1\rangle\}$. However, measurements in other bases are also possible by applying appropriate state rotation gates before the measurement. In this section, we present a qubit state readout mechanism that relies on a qubit-dependent shift in the oscillator's resonance frequency. This approach is similar to how the state of a superconducting qubit is measured by tracking the resonance frequency of a coupled waveguide resonator~\cite{Blais:2004,Wallraff:2004}. We compare the resonance frequency shift for both bare and dressed qubits and analyze how nonzero thermal occupation influences each configuration.

In the dispersive regime $|\Delta_R|\gg g$, the resonance frequency of the harmonic oscillator gets shifted by $\pm\frac{g^2}{\Delta_R}$ depending on the qubit's state~\cite{Blais:2004,Gambetta:2006}, see Eq.~\eqref{eq:frequency_shift}. The resonance frequency manifests as a maximum in the noise power spectral density, Eq.~\eqref{eq:Sx}. The shift of the maximum can be detected, for instance, in an optomechanical quantum bus, through optical measurements, and pulsed measurements allow for the detection of short-lived shifts in the mechanical frequency~\cite{Lyyra:2024,Vanner:2011,Vanner:2013,Muhonen:2019}. The resonance frequency shift is generally measurable if it is greater than the linewidth.

In this section, we analyze the frequency- and time-domain response of a qubit coupled to a harmonic oscillator. We solve the QME~\eqref{eq:QME} for a bare qubit near-resonant with the harmonic oscillator, described by Hamiltonian~\eqref{eq:bare qubit}, and a dressed qubit (with off-resonant bare transition frequency), described by Hamiltonian~\eqref{eq:dressed Hamiltonian}. The codes used to solve the equation are available at~\cite{programs}. The temperature $T$ of the heat bath determines the thermal occupations $n_{\mathrm{th},q}$ and $n_{\mathrm{th},h}$ in the Lindblad operators, and the initial state of the harmonic oscillator, which we assume to be in thermal equilibrium with the heat bath so that $\rho_h(0)=Z^{-1}e^{-\omega_h\hat{a}^\dagger\hat{a}/T}$, where $Z$ is the partition function of the harmonic oscillator. The qubit is initialized either in the ground ($|0\rangle$) or the excited ($|1\rangle$) state. The noise power spectral density introduced in Eq.~\eqref{eq:Sx} is written in the Heisenberg picture, while the QME formalism is based on the Schrödinger picture. Using the quantum regression theorem~\cite{Gardiner_book,Lax:1963}, the two-time correlation function in Eq.~\eqref{eq:Sx} can be rewritten as
\begin{equation}\label{eq:correlator_schrodinger}
    \langle\{\hat{x}_\mathrm{H}(t),\hat{x}_\mathrm{H}(0)\}\rangle=\Tr{\hat{x}_\mathrm{S}\hat{\tilde{\rho}}(t)}\; ,
\end{equation}
where $\hat{x}_\mathrm{S}$ and $\hat{x}_\mathrm{H}$ are the position operator in the Schrödinger and Heisenberg pictures, respectively. Equation~\eqref{eq:correlator_schrodinger} can be interpreted as the expectation value of $\hat{x}$ for an unnormalized state described by the matrix $\hat{\tilde{\rho}}=\{\hat{x},\hat{\rho}\}$. Solving the QME for $\hat{\tilde{\rho}}(t)$, with the initial condition $\hat{\tilde{\rho}}(0)=\{\hat{x}_\mathrm{S},\hat{\rho}(0)\}$, we evaluate $\eqref{eq:correlator_schrodinger}$ and perform a fast Fourier transform to compute $S(\omega)$ [Eq.~\eqref{eq:Sx}].

In Fig.~\ref{fig:spring_effect} we show the spectral density of the $\hat{x}$ quadrature [Eq.~\eqref{eq:Sx}] for a dressed qubit [Fig.~\ref{fig:spring_effect}(a)] and a bare qubit [Fig.~\ref{fig:spring_effect}(b)]. In order to compute the Fourier transform in Eq.~\eqref{eq:Sx}, we have considered a finite integration time. Since the system is damped, it eventually reaches a steady state, which at long integration times dominates the integral for the Fourier transform and thus erases any information from the initial state. This cutoff represents the measurement time in an experiment, and it defines the frequency resolution as $\Delta\omega=2\pi/t_{\mathrm{cut}}$. The resonance frequency shift lies in the $\omega-\omega_h\in[-g^2/|\Delta_R|,g^2/|\Delta_R|]$ range, so the frequency resolution should be high enough to distinguish positive and negative shifts. For the numerical computations, we have chosen a cutoff time of $t_{\mathrm{cut}}=40\pi|\Delta_R|/g^2$, which provides sufficient frequency resolution to accurately determine the position of the resonance peak.

For the dressed qubit, we assume that the driving field detuning is negligible compared to the coupling constant and the damping rates and set $\Delta=0$. We consider a Rabi detuning, i.e., the detuning between the drive amplitude and the harmonic oscillator $\Delta_R=\Omega_R-\omega_h$, of $\Delta_R=0.05\omega_h$. The sign of $\Delta_R$ determines the resonance frequency shift direction for the ground and excited states of the qubit, while the magnitude of $\Delta_R$ quantifies the detuning between the dressed qubit and the harmonic oscillator frequency. We choose a coupling constant $g=5 \cdot 10^{-3}\omega_h$, and the qubit and harmonic oscillator damping rates are $\gamma_q=10^{-4}\omega_h$ and $\gamma_h=10^{-4}\omega_h$, respectively. A small coupling constant $g \ll |\Delta_R|$ guarantees that the system remains in the dispersive regime, necessary for the implementation of the two-qubit gate [see Sec.~\ref{sec:gate}]. We have chosen damping rates of the order of $\gamma_q,\gamma_h\sim\frac{g^2}{|\Delta_R|}$ to study thermal effects. In practice, damping rates should be as low as possible, since their effect is generally detrimental. However, in the case of detecting the resonance frequency from the noise power spectral density, a higher damping rate increases the resonance peak bandwidth, making detection easier.

As shown in Fig.~\ref{fig:spring_effect}(a), in the absence of thermal occupation of the oscillator (blue line), the resonance frequency shift for the dressed qubit is given approximately by $\pm\frac{g^2}{\Delta_R}$ [Eq.~\eqref{eq:frequency_shift}], with the upper and lower signs corresponding to the excited and ground states, respectively. The resonance frequency shift is slightly suppressed at higher temperatures, but the resonance peak height increases, making it easier to detect the resonance frequency. $S(\omega)$ shows a small kink at the frequency corresponding to the opposite state $\mp\frac{g^2}{\Delta_R}$. This feature can be explained by studying the time-domain response. In Fig.~\ref{fig:spring_effect}(c) we show the time evolution of a dressed qubit with initial state $|1\rangle$. The system begins at the excited state with $\langle\hat{\sigma}_x\rangle=1$ (see Hamiltonian~\eqref{eq:dressed Hamiltonian}), and evolves into a mixed state due to the damping~\cite{atom-photon-interactions-book}. As shown in Appendix~\ref{sec:stationary_state}, in the steady state, the weight of the ground state is higher than that of the excited state, so that $\langle\hat{\sigma}_x\rangle<0$. The dashed line in Fig.~\ref{fig:spring_effect}(c) indicates the value of $\langle\hat{\sigma}_x\rangle$ in the steady state, given by Eq.~\eqref{eq:stationary_state}. The contribution of the $|0\rangle$ state to the noise power spectral density leads to the kinks at $\omega=\omega_h-\frac{g^2}{\Delta_R}$.

In Figs.~\ref{fig:spring_effect}(b) and (d), we consider a bare qubit with a detuning equivalent to that of the dressed qubit and set $\omega_q=1.05\omega_h$. The logical states of the bare qubit are given by the eigenstates of $\hat{\sigma}_z$ [see Hamiltonian~\eqref{eq:bare qubit}]. Unlike the dressed qubit, at high temperatures the position of the resonance frequency peak gets sizably shifted to lower values, going below the bare resonance frequency of the oscillator for $n_{\mathrm{th},h}\sim 4$ (yellow line). Moreover, the kink at the opposite frequency becomes very relevant at nonzero thermal occupation, so that the initial state of the qubit cannot be determined from the noise power spectral density for $n_{\mathrm{th},h}\gtrsim 4$. As shown in Fig.~\ref{fig:spring_effect}(d), a qubit with initial state $|1\rangle$ rapidly decays to the ground state ($\langle\hat{\sigma}_z\rangle=-1$), leading to a sizable peak at $\omega=\omega_h-\frac{g^2}{\Delta_R}$. For the bare qubit, $\omega_q$ needs to be nearly resonant with the harmonic oscillator $\omega_q \sim \omega_h$, so the bare qubit's quality factor $Q_q=\omega_q/\gamma_q$ is smaller than the dressed qubit's one, where $\omega_q \gg \omega_h$. Another drawback of using the bare qubit in resonance with the low-frequency harmonic oscillator is that since $\omega_q\sim\omega_h$, the qubit thermal occupation $n_{\mathrm{th},q} \sim n_{\mathrm{th},h}$ is not negligible, leading to a stronger damping effect. Therefore, dressing the qubit does not only provide a mechanism to control the qubit-harmonic oscillator detuning, but it also leads to a lower effective temperature for the qubit. In summary, the dressed qubit is more suitable for the readout mechanism based on the spectral density of the oscillator, since $n_{\mathrm{th},q}$, and consequently the effective damping rate $(n_{\mathrm{th},q}+1)\gamma_q$, are lower for the dressed qubit than in a bare qubit nearly resonant with the oscillator.

\section{Implementation of a two-qubit gate}
\label{sec:gate}

\begin{figure*}
    \centering
    \includegraphics[width=0.9\textwidth]{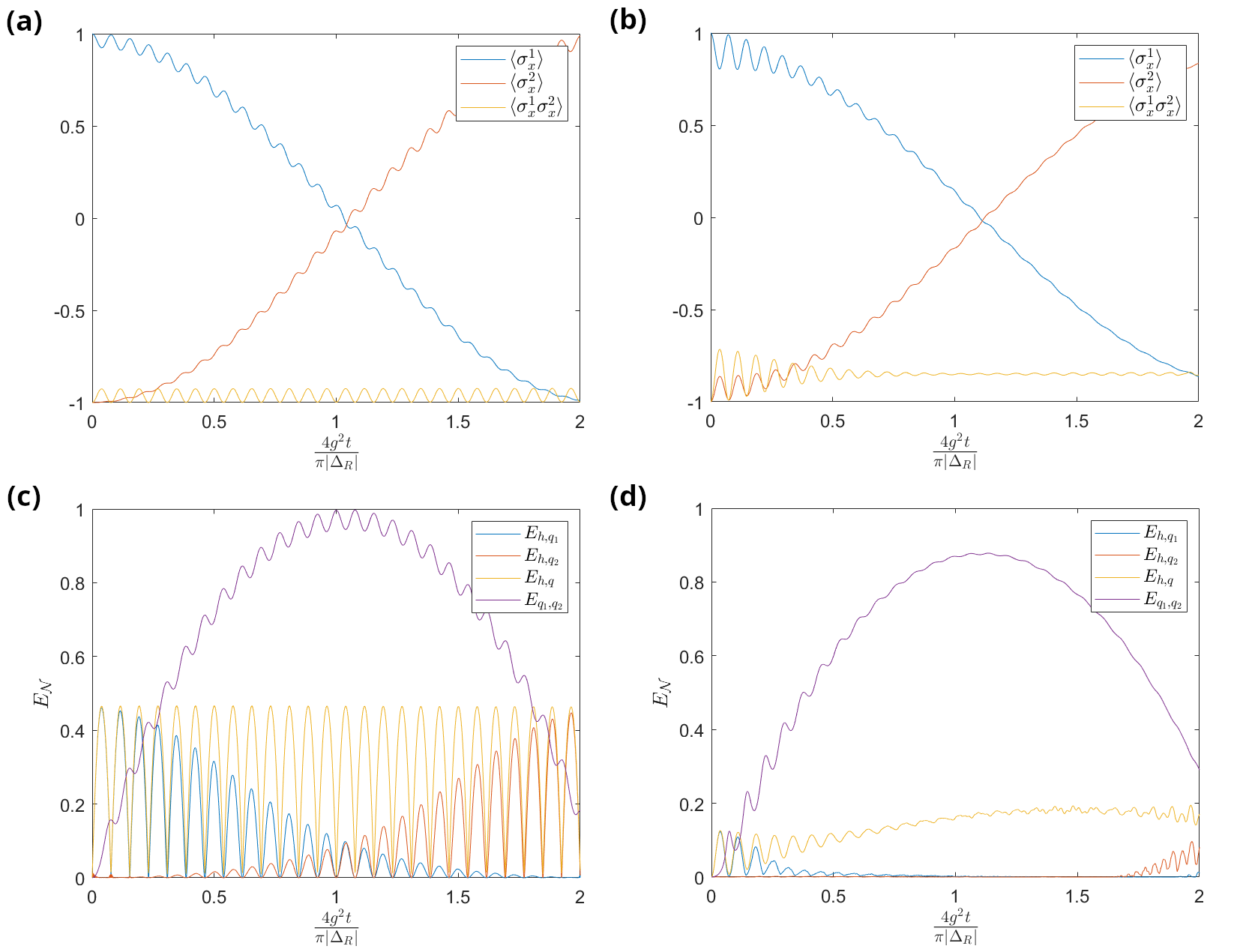}
    \caption{Evolution of the qubit expectation values for an initial state $|10\rangle$ for thermal excitation amplitudes (a) $n_{\mathrm{th},h}=0$ and (b) $n_{\mathrm{th},h}=2$. Entanglement between the different subsystems as a function of time for (c) $n_{\mathrm{th},h}=0$ and (d) $n_{\mathrm{th},h}=2$. All plots use the parameters $\Delta=0$, $\Delta_R=5 \cdot 10^{-2}\omega_h$, $g=5 \cdot 10^{-3}\omega_h$, $\gamma_q=10^{-6}\omega_h$ and $\gamma_h=10^{-6}\omega_h$.}
    \label{fig:entanglement}
\end{figure*}

In the previous section, we characterize the coupling between the qubit and the harmonic oscillator, highlighting the benefits of using a driven qubit for state readout. In the following, we consider two driven qubits coupled to the same harmonic oscillator, as illustrated in Fig.~\ref{fig:sketch}(b). We study the implementation of a two-qubit gate, where the qubit-qubit interaction is mediated by the harmonic oscillator. The Hamiltonian for a two-qubit system can be obtained by a direct generalization of Hamiltonian~\eqref{eq:dressed Hamiltonian}
\begin{equation}\label{eq:dressed Hamiltonian 2}
    \hat{H}_\mathrm{dress}=\sum_j\left(-\frac{\Delta_j}{2}\hat{\sigma}_z^j+\frac{\Omega_{R,j}}{2}\hat{\sigma}_x^j\right)+\omega_h\hat{a}^\dagger \hat{a}-\sum_j g_j\hat{\sigma}_z^j(\hat{a}+\hat{a}^\dagger)\;.
\end{equation}
This formalism can be readily extended to multiqubit systems, where any two qubits can be selectively coupled through dressing. Furthermore, individual qubit states can be measured by dressing a single qubit and following the procedure described in Sec.~\ref{sec:characterization}.

Within the rotating wave approximation, the coupling between the qubit and the harmonic oscillator gives rise to an effective qubit-qubit interaction~\cite{Zheng:2000,Blais:2004,Blais:2007,Liu:2006,Srinivasa:2024}, see Appendix~\ref{sec:iSWAP}. The interaction time between the qubits can be controlled by tuning their drive amplitudes $\Omega_{R,j}(t)$ close to resonance with the harmonic oscillator. Assuming equal drive amplitudes and coupling strengths for both qubits, the unitary evolution of the system for an interaction time of $t_\mathrm{int}=\frac{\pi|\Delta_R|}{4g^2}$ corresponds to a $\sqrt{i\mathrm{SWAP}}$-gate operation, up to one-qubit transformations,
\begin{equation}\label{eq:sqrt(iSWAP)}
    \sqrt{\mp i\mathrm{SWAP}}=
    \begin{pmatrix}
        1 & 0 & 0 & 0\\
        0 & 1/\sqrt{2} & \mp i/\sqrt{2} & 0\\
        0 & \mp i/\sqrt{2} & 1/\sqrt{2} & 0\\
        0 & 0 & 0 & 1
    \end{pmatrix}\; ,
\end{equation}
where the upper and lower signs correspond to a positive and negative Rabi detuning. The $\sqrt{i\mathrm{SWAP}}$-gate generates entanglement between the qubits if they are in opposite logical states. For instance, it maps the separable state $|10\rangle$ into the maximally entangled state $(|10\rangle+i|01\rangle)/\sqrt{2}$. In Appendix~\ref{sec:Bell_states} we explain how to create Bell states in driven qubits~\cite{Liu:2006}. In the following, we study how the qubit-harmonic oscillator coupling enables the qubit-qubit entanglement generation, and analyze the gate fidelity to determine a suitable range of the system parameters for a practical implementation.

\subsection{Qubit entanglement}

Universal quantum computation requires the capacity to generate any pure state from an arbitrary initial state. Typically, qubits are initialized in a separable state, which can be manipulated into an entangled state through a sequence of quantum gates. In this section, we analyze the capacity to generate a fully entangled two-qubit state, since this provides a benchmark for the gate's fidelity. The qubit-qubit coupling is mediated by a harmonic oscillator, which may become entangled with the qubits during the process. The main goal of this section is to determine whether the qubit-harmonic oscillator system results in qubit-qubit entanglement, or if these are independent processes, and to determine how thermal effects hinder the generation of a fully entangled state. Entanglement in pure bipartite systems is typically measured by the von Neumann entropy of the reduced density matrices. However, for mixed or multipartite systems, the von Neumann entropy is not always a suitable measure, as a classically correlated mixed state can have nonzero von Neumann entropy even though it is not entangled. Several alternative measures of entropy exist for mixed systems~\cite{Bennett:1996A,Bennett:1996B}. Here, we use logarithmic negativity~\cite{Vidal:2002,Plenio:2005} as a measure of entanglement. Given two subsystems, $A$ and $B$, logarithmic negativity is defined as
\begin{equation}\label{eq:negativity}
    E_\mathcal{N}(\hat{\rho}) \equiv \log_2{||\hat{\rho}^{\mathrm{T}_A}||}\; ,
\end{equation}
where $\hat{\rho}^{\mathrm{T}_A}$ is the partial transpose with respect to subsystem $A$ and $||\hat{X}|| = \mathrm{Tr} \sqrt{\hat{X}^\dagger \hat{X}}$ is the trace norm. Unlike the entropy of entanglement, logarithmic negativity is entanglement monotone~\cite{Vidal:2000,Plenio:2005} both for pure and mixed states. For 2-level systems such as qubits, logarithmic negativity is normalized such that $E_\mathcal{N}=1$ for a maximally entangled state.

We solve the quantum master equation~\eqref{eq:QME} for the system Hamiltonian~\eqref{eq:dressed Hamiltonian 2}~\cite{programs}. To compute the entanglement between the different subsystems, we perform partial traces over $\hat{\rho}(t)$ to obtain the relevant reduced density matrix. In Fig.~\ref{fig:entanglement}(a), we show the evolution of the two-qubit states for two consecutive applications ($0 \leq t \leq 2t_\mathrm{int}$) of the $\sqrt{-i\mathrm{SWAP}}$ gate, with initial state $|10\rangle$ and thermal occupation $n_{\mathrm{th},h}=0$. Ideally, the system should evolve into the $(|10\rangle-i|01\rangle)/\sqrt{2}$ state at $t=t_\mathrm{int}$, and to the $-i|01\rangle$ state at $t=2t_\mathrm{int}$, swapping the states of the qubits up to a phase factor. On top of the swapping of the qubit occupation numbers $\hat{\sigma}_{x,i}$, there are small oscillations with frequency $\Delta_R+4g^2/\Delta_R$ and an amplitude of the order of $2g^2/\Delta_R^2$. The cross-correlation $\langle\hat{\sigma}_x^1\hat{\sigma}_x^2\rangle$ oscillates close to -1, since the qubits ideally remain in opposite states. The small oscillations arise due to deviation of the parameters from the dispersive regime, i.e., by retaining higher-order terms in the $g \ll |\Delta_R|$ expansion carried out in Appendix~\ref{sec:iSWAP}. In the plot, we have chosen values of $g$ and $\Delta_R$ satisfying $(\Delta_R+4g^2/\Delta_R)t_\mathrm{int} \in 2\pi\mathbb{N}$, so that $\langle\hat{\sigma}_x^1\hat{\sigma}_x^2\rangle$ is minimal at $t=t_\mathrm{int}$, thus maximizing the gate fidelity $\mathcal{F}$. Nonetheless, the crossing point of the occupation numbers $\langle\hat{\sigma}_x^1\rangle=\langle\hat{\sigma}_x^2\rangle=0$ occurs at a time slightly higher than $t=t_\mathrm{int}$. A finer tuning of $\Delta_R$ and $g$ could simultaneously optimize the qubit expectation values to enhance the fidelity.

In Fig.~\ref{fig:entanglement}(b), we show the evolution of the qubit states for $n_{\mathrm{th},h}=2$. Higher temperature leads to enhanced damping [see Eq.~\eqref{eq:Lindblad}], so the small oscillations fade out as time progresses. As a result, $\langle\hat{\sigma}_x^1\hat{\sigma}_x^2\rangle$ may not reach a value of $\langle\hat{\sigma}_x^1\hat{\sigma}_x^2\rangle=-1$ at $t=t_\mathrm{int}$. Moreover, state swapping dynamics slow down, delaying the crossing time of $\langle\hat{\sigma}_{x,i}\rangle$.

In Fig.~\ref{fig:entanglement}(c-d) we show the entanglement, quantified by the logarithmic negativity~\eqref{eq:negativity}, between the different subsystems, namely the harmonic oscillator ($h$), the subsystem formed by the two qubits ($q$), and each qubit ($q_1$ and $q_2$). The system begins in a separable state and evolves into a state with maximally entangled qubits at $t=t_\mathrm{int}$ (purple line). The harmonic oscillator-qubit entanglement $E_{h,q}$ (yellow line) oscillates periodically with frequency $\Delta_R+4g^2/\Delta_R$ in antiphase with the $q_1-q_2$ entanglement. This periodicity can be used to tune the values of $g$ and $|\Delta_R|$ to maximize entanglement, and therefore, the fidelity of the gate. The amplitudes of the oscillations of $E_{h,q_1}$ (blue line) and $E_{h,q_2}$ (red line) are modulated by their respective qubit's state. Since the qubits are in near resonance with the harmonic oscillator, one may simplify the coupling term in Eq.~\eqref{eq:dressed Hamiltonian 2} by applying the rotating wave approximation. Within the rotating wave approximation, doubly rotating terms are disregarded, so that the remaining terms describe energy quanta exchange between the qubit and the harmonic oscillator ($\hat{\sigma}_{-,x}^j\hat{a}^\dagger$ and $\hat{\sigma}_{+,x}^j\hat{a})$. Here, $\hat{\sigma}_{\pm,x}$ are the raising and lowering operators in the dressed qubit basis. Since at $T=0$ the harmonic oscillator is initialized in the ground state, at $t=0$ it may only couple to the first qubit, which starts in the excited state $|1\rangle$. The situation is reversed at $t=2t_\mathrm{int}$.

At a nonvanishing temperature, Fig.~\ref{fig:entanglement}(d), in agreement with the results of Fig.~\ref{fig:entanglement}(b), the oscillations of the entanglement get suppressed as time progresses. However, close to $t=2t_\mathrm{int}$, higher order harmonics become relevant. While entanglement is not a conserved quantity, the general trend in Figs.~\ref{fig:entanglement}(c-d) is the pumping of the $h-q_1$ entanglement into the $q_1-q_2$ entanglement for $t \lesssim t_\mathrm{int}$, and the pumping of $q_1-q_2$ to $h-q_2$ entanglement for $t_\mathrm{int} \lesssim t \lesssim 2t_\mathrm{int}$. Therefore, the suppression of the $h-q_j$ entanglements at high temperatures lowers the ability of the quantum gate to generate a maximally entangled $q_1-q_2$ state.

\subsection{Fidelity of the two-qubit gate}\label{sec:fidelity}

\begin{figure*}
    \centering
    \includegraphics[width=0.99\textwidth]{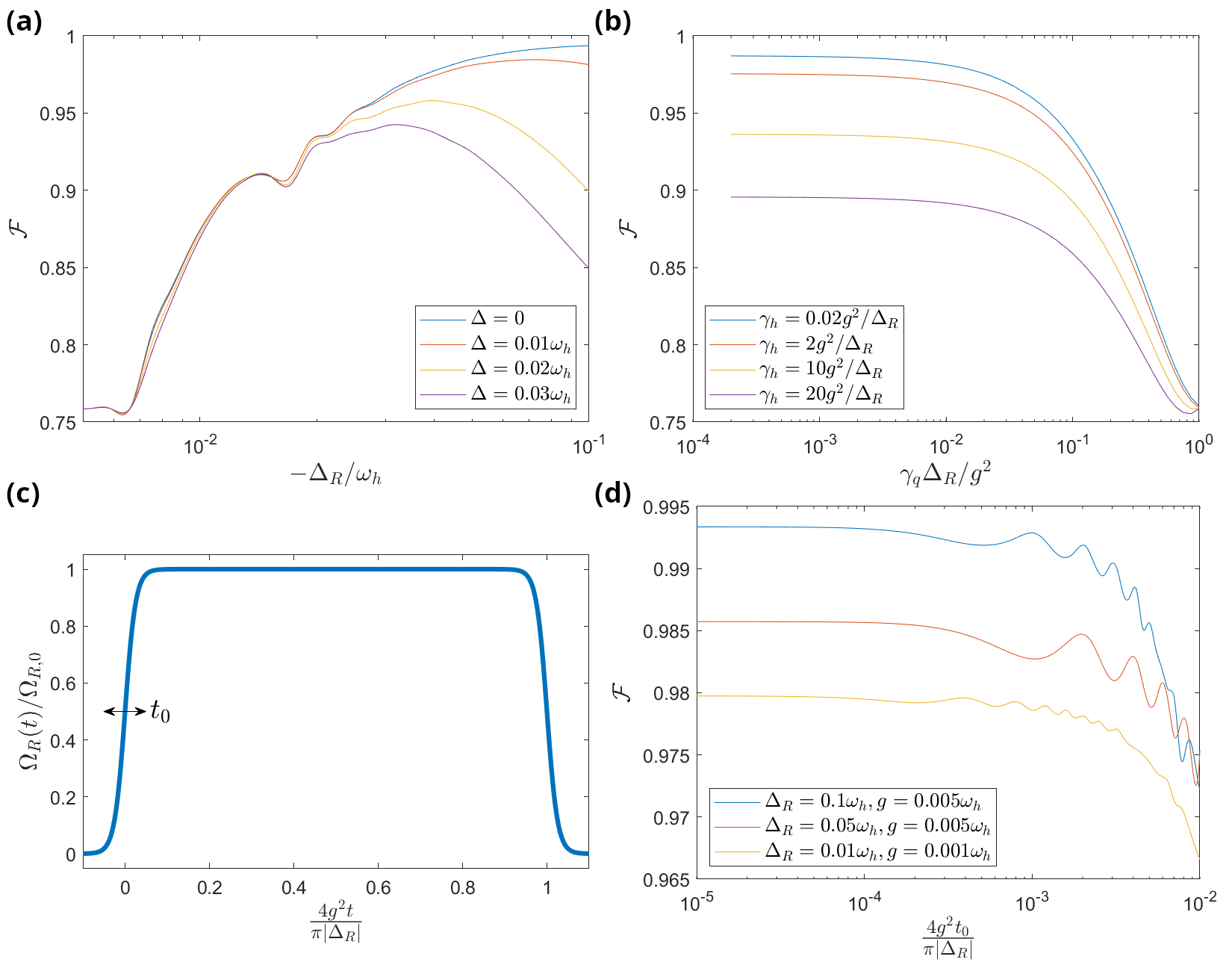}
    \caption{(a) Fidelity of the $\sqrt{i\mathrm{SWAP}}$-gate as a function of the driving field detuning and the Rabi detuning. The parameters used are $n_{\mathrm{th},h}=0$, $g=5 \cdot 10^{-3}\omega_h$, $\gamma_q=10^{-6}\omega_h$ and $\gamma_h=10^{-6}\omega_h$. (b) Fidelity of the quantum gate as a function of the qubit and harmonic oscillator dampings. The parameters used are $n_{\mathrm{th},h}=0$, $\Delta=0$, $\Delta_R=5 \cdot 10^{-2}\omega_h$ and $g=5 \cdot 10^{-3}\omega_h$. (c) Depiction of the dressed qubit transition frequency for a nonvanishing dressing rise time, and (d) fidelity of the $\sqrt{i\mathrm{SWAP}}$-gate as a function of the rise time. All plots use the parameters $n_{\mathrm{th},h}=0$, $\Delta=0$, $\gamma_q=10^{-6}\omega_h$ and $\gamma_h=10^{-6}\omega_h$.}
    \label{fig:fidelity}
\end{figure*}

In the previous section we study how temperature affects the qubit-qubit entanglement generation, which is crucial for universal quantum computation. The thermal occupation of the system is determined by the temperature of its environment. As a rule of thumb, higher temperatures are detrimental to coherence. The system parameters, such as characteristic frequencies and coupling constants, are also critical in achieving high-fidelity quantum computing. Typically, these parameters are fixed by external electric or magnetic fields, materials choices, or the physical realization of the quantum register. In the following, we study how different parameters of the model affect the fidelity of the $\sqrt{i\mathrm{SWAP}}$-gate.

Fidelity $\mathcal{F}$ is a measure of how closely a quantum gate's implementation matches the ideal behavior ($\mathcal{F}=1$). For pure states, it is given by the inner product of the real and ideal final states, averaged over all initial pure states. However, thermal fluctuations and damping drive the system into a mixed state, so the definition of the gate fidelity takes a more general form:
\begin{equation}\label{F_Schrodinger}
    \mathcal{F}(t)=\overline{\langle\varphi|\hat{\rho}_q(t)|\varphi\rangle}\; ,
\end{equation}
where $\hat{\rho}_q(t)$ is the two-qubit reduced density matrix, $|\varphi\rangle$ is the expected output for a given initial state, and the overline denotes averaging over all initial states. We numerically evaluate the fidelity of the quantum gate by solving the QME~\eqref{eq:QME} over an ensemble of initial states homogeneously distributed over the Bloch sphere, and averaging the fidelity according to Eq.~\eqref{F_Schrodinger}~\cite{programs}.

In Fig.~\ref{fig:fidelity}(a) we show how the driving field detuning $\Delta$ and the Rabi detuning $\Delta_R$ limit the fidelity. Fidelity decays monotonically with an increasing $\Delta$. $\Delta$ and $\Omega_R$ determine the qubit's eigenstates [see Eq.~\eqref{eq:dressed Hamiltonian}], so the dressed qubit's eigenbasis differs from the computational basis for a nonvanishing $\Delta$. The qubit-harmonic oscillator resonance condition is given by $\omega_h=\sqrt{\Omega_R^2+\Delta^2}$~\cite{Srinivasa:2024}. For $\Delta=0$, the qubit-harmonic oscillator detuning is symmetric in $\Delta_R$, but a nonvanishing $\Delta$ leads to a higher detuning for $\Delta_R=\Omega_R-\omega_h>0$. A perfect tuning of the driving field is not realistic in practical implementations, so a negative $\Delta_R$ is preferable for qubit-harmonic oscillator resonance. As shown in Ref.~\cite{Lyyra:2024}, for $\Delta_R<0$ a nonzero $\Delta$ enhances the resonance frequency shift. Therefore, in Fig.~\ref{fig:fidelity}(a) we focus on the $\Delta_R<0$ regime. The fidelity suppression due to a nonzero $\Delta$ is most significant for higher Rabi detunings $|\Delta_R|$.

Fidelity is non-monotonous in $\Delta_R$. On the one hand, the rotating wave approximation necessary for building the harmonic oscillator-mediated $\sqrt{i\mathrm{SWAP}}$-gate [see Appendix~\ref{sec:spring effect}], requires a small Rabi detuning $|\Delta_R|\ll\omega_h$. On the other hand, the dispersive regime is valid for $g\ll\Delta_R$. Focusing on $\Delta=0$ (blue line), the rotating wave approximation is well satisfied for Rabi detunings as large as $|\Delta_R| \approx 0.1\omega_h$, but the fidelity has a significant drop for $|\Delta_R|/g < 10$ due to deviations from the dispersive regime. Regarding the driving field detuning, a nonvanishing $\Delta$ changes the qubits' eigenbasis. Therefore, a high-fidelity gate implementation requires that $\Delta\ll\Omega_R,\omega_h$. Based on the results of Fig.~\ref{fig:fidelity}(a), an appropriate range of values for the detunings is $\Delta<0.01\omega_h$ and $10g<|\Delta_R| \lesssim 0.1\omega_h$. For a given $\Delta$, $|\Delta_R|$ may be calibrated to maximize $\mathcal{F}$.

In Fig.~\ref{fig:fidelity}(b) we study the dependence of the gate fidelity on the damping rates. The damping rates quantify the coupling of the system to the environment, so higher damping rates will always amount to lower $\mathcal{F}$. Although the dressed qubit and harmonic oscillator characteristic frequencies are approximately the same $\Omega_R\sim\omega_h$, the qubit damping rate has a more critical impact on $\mathcal{F}$, since it directly acts on the qubit degrees of freedoms. The inverse qubit damping rate defines the qubit decoherence time, so the damping rate should satisfy $\gamma_qt_\mathrm{int}\ll 1$ to ensure that the qubit remains coherent throughout the application of the quantum gate. Numerical results show that the fidelity decreases significantly for values exceeding $\gamma_q \gtrsim 10^{-2}g^2/\Delta_R$, whereas the harmonic oscillator rate can be as high as $\gamma_h=2g^2/\Delta_R$ while producing the same fidelity loss.

Until now we have assumed that the qubit-harmonic oscillator interaction time $t_\mathrm{int}$, determined by the dressing of the qubit, is perfectly controlled. However, in practice, bringing the amplitude of the driving field $\Omega_R$ into resonance with the harmonic oscillator frequency $\omega_h$ requires a finite amount of time $t_0$. Imperfect control of the qubit–qubit interaction time leads to a reduction in the fidelity of the two-qubit gate. We consider a dressing amplitude described by
\begin{equation}
    \Omega_R(t)=\frac{\Omega_{R,0}}{4}(1+\tanh(2t/t_0))(1-\tanh(2(t-t_\mathrm{int})/t_0))\; ,
\end{equation} as depicted in Fig.~\ref{fig:fidelity}(c). In Fig.~\ref{fig:fidelity}(d) we show how the gate fidelity depends on the rise time. Based on the previous analysis of the system parameters, we select values of the detunings and coupling constant suitable for quantum computation. Modern quantum computers achieve two-qubit gate fidelities of approximately $\mathcal{F} \sim 99\%$~\cite{Abdurakhimov:2024,McKay:2024,Barends:2014}. A minimum fidelity of $\mathcal{F} \sim 99.99\%$~\cite{Knill:2005,Sutcliffe:2025}, combined with quantum error correction protocols, is expected for practical quantum computing. The general trend shows that gate fidelity decreases with slower rise times, but there is also an oscillatory behavior that may be exploited to enhance $\mathcal{F}$. Given the practical limitations on the rise time, one may calibrate the rise time to maximize the fidelity.

Besides the nonidealities of the two-qubit gate, the cross-talk between the qubits while the dressing is off also affects fidelity. Following the same procedure as in Appendix~\ref{sec:iSWAP} for undressed qubits ($\Omega_R=0$), where the rotating wave approximation is not valid due to the qubit-harmonic oscillator detuning, the effective qubit-qubit interaction is described by the Hamiltonian $\hat{H}=4\frac{g^2}{\omega_h}\hat{\sigma}_z^1\hat{\sigma}_z^2$. For the optimal parameters derived in this section, the residual interaction is of order of $g^2/\omega_h\sim 10^{-5}\omega_h$, making it comparable to decoherence.

\section{Practical realization}
\label{sec:realization}
Most present-day quantum computer realizations rely on qubits coupled via electromagnetic resonators or with other qubits connected to each other electromagnetically~\cite{Wallraff:2004}. Such interqubit coupling requires that the qubit states are discernible based on their different charge susceptibilities. Those qubits are inherently susceptible to offset charge fluctuations, even if the effective coupling to those fluctuations can be minimized as done, for example, in superconducting transmon qubits~\cite{Houck:2009}. The longest single-qubit coherence times are therefore obtained with qubits having no coupling to charge, such as spin qubits~\cite{muhonen2014}. Then, the problem is to find a suitable mediator for coupling such qubits while maintaining their good coherence. One possible choice is based on utilizing mechanical vibrations in a mechanical resonator, coupling to the qubit states via any mechanism where the resonator's motion shifts the (Zeeman) spin splitting, such as the qubit strain-induced effects~\cite{mansir2018}, or a motion-modulated external magnetic field. The latter mechanism hence requires a spatial magnetic field gradient, and the coupling strength is proportional to the amount of change in the magnetic field within the zero-point vibration amplitude. 

As many high-quality mechanical resonators have resonant frequencies within the few MHz range, but bringing the qubits to their ground state in dilution refrigerator conditions (temperatures of the order of tens of mK) require at least GHz range qubit frequencies, the dressed qubit approach considered in this manuscript is the only feasible approach to bring these two systems into resonance. Moreover, the mechanical resonators can also couple to electromagnetic fields via the non-linear optomechanical coupling. This coupling imprints the mechanical spectral density in the noise spectral density of the phase quadrature of the electromagnetic field. Thus, it allows for the readout of the qubit state via the measurement of the electromagnetic field's noise power spectral density. In addition, it allows for an optomechanical cooling of the mechanical resonator close to its ground state~\cite{Park:2009}, diminishing the thermal noise ($n_{\mathrm{th},h}$) affecting the harmonic oscillator. 

Although the scheme described in this manuscript is rather generic, a scheme for realizing it in the context of bismuth or phosphorous donor-based silicon spin qubits coupled to an optomechanical resonator is discussed in Ref.~\cite{Lyyra:2024}.

\section{Conclusions}
\label{sec:conclusions}
In this work, we thoroughly analyze the use of driven qubits coupled to a harmonic oscillator to implement a $\sqrt{i\mathrm{SWAP}}$-gate. In many experimental realizations, the characteristic frequencies of the bare qubits and the bus differ by several orders of magnitude. In such cases, dressing the qubits via a driving field provides a mechanism for selectively coupling the qubits to the bus. First, we investigate thermal effects on a qubit readout mechanism based on the measurement of the resonance frequency of the harmonic oscillator~\cite{Vanner:2011,Vanner:2013,Muhonen:2019}, comparing the results for bare and dressed qubits. We show that the qubit state readout mechanism is more robust for the dressed qubit, as the steady state in the former generally exhibits information on both initial qubit states, and it is sufficient to fix the measurement time by observing the behavior of the different amplitudes of the peaks. In contrast, for bare qubits, the resonance peak describing the steady state may mask the peak associated with the initial transient state. Moreover, dressed qubits tolerate thermal occupations better. We have also studied how the qubit-harmonic oscillator coupling can be leveraged to implement a two-qubit gate. We have analyzed the influence of various model parameters on the gate fidelity, identifying an optimal parameter regime for quantum computation.

This setup may achieve two-qubit quantum gate fidelities on par with those of other physical platforms~\cite{Tanttu:2024,Huang:2024}, but the constraints of the applicability of the rotating wave approximation and the dispersive regime impair the capacity to obtain a fidelity compatible with practical quantum computing $\mathcal{F} \sim 99.99\%$. The scheme is quite versatile and can be applied to any type of qubits coupled to resonators. For example, in the case of nanomechanical resonators, this approach could enable compact qubit storage, paving the way for chips with a higher qubit count. Combined with the fact that qubit dressing offers a means to control the qubit-harmonic oscillator interaction time and induces a lower effective qubit temperature, this scheme presents a promising alternative to existing quantum computing architectures. Future works may study how crosstalk with undressed qubits may affect the fidelity of the state readout and the two-qubit gate in multiqubit systems.

One key feature of using dressed qubits is the balanced steady state. A bare qubit is strongly affected by environmental damping, leading to relaxation into its ground state $|0\rangle$. In contrast, a dressed qubit relaxes into a mixed state, symmetrizing the $|0\rangle$ and $|1\rangle$ logical states. The state readout time may exceed the qubit's decoherence time, so having a mixed steady state helps prevent a qubit flip due to thermal relaxation, thus improving the efficiency of initial state determination. However, most quantum algorithms require pure states, making the steady state of a dressed qubit unsuitable for qubit initialization. A potential solution is to suppress the dressing amplitude $\Omega_R$ and apply a Zeeman field along the $z$-direction, enabling the qubit to decay into its bare ground state via thermal relaxation.

\renewcommand\acknowledgmentsname{Acknowledgements}
\begin{acknowledgments}
This work was funded by the Research Council of Finland (project no. 354735, 321416 and 359240) and the European Research Executive Agency (grant agreement No 101202316). We acknowledge grants of computer capacity from the Finnish Grid and Cloud Infrastructure (persistent identifier urn:nbn:fi:research-infras-2016072533). This project has received funding from the European Research Council (ERC) under the European Union’s Horizon 2020 research and innovation programme (Grant Agreement No. 852428).
\end{acknowledgments}

\appendix
\numberwithin{equation}{section}
\renewcommand{\thesubsection}{\arabic{subsection}}

\section{Resonance frequency shift}\label{sec:spring effect}
If the drive amplitude (Rabi frequency) $\Omega_R$ is close to resonance with the harmonic oscillator, we may apply the rotating wave approximation to the qubit-harmonic oscillator to simplify the coupling term. Working in the dressed qubit basis, the Hamiltonian becomes:
\begin{equation}\label{eq:RWA Hamiltonian}
    \hat{H}_\mathrm{dress}=\frac{\Omega_R}{2}\hat{\sigma}'_z+\omega_h\hat{a}^\dagger \hat{a}+g(\hat{\sigma}'_-\hat{a}^\dagger+\hat{\sigma}'_+\hat{a})\; ,
\end{equation}
where $\hat{\sigma}'_z=\sigma_x$ and $\hat{\sigma}'_\pm=(-\sigma_z\pm i\sigma_y)/2$ are the Pauli operators in the dressed qubit basis. The coupling term in~\eqref{eq:RWA Hamiltonian} conserves the total number of quanta of the system, and it can be exactly diagonalized in the $\{|n,1\rangle,|n+1,0\rangle\}$ subspaces~\cite{introductory-quantum-optics}.

Hamiltonian~\eqref{eq:RWA Hamiltonian} can be further simplified in the dispersive regime, where the detuning between the dressed qubit and the harmonic oscillator $\Delta_R=\Omega_R-\omega_h$ is large compared to the coupling strength $|\Delta_R|\gg g$ and $\langle \hat{a}^\dagger\hat{a}\rangle\ll\Delta_R^2/(4g^2)$~\cite{Blais:2004,Boissonneault:2009}. Applying a Schrieffer-Wolff transformation~\cite{Schrieffer–Wolff} of the form $\hat{U}=e^{g/\Delta_R(\hat{a}\hat{\sigma}'_+ - \hat{a}^\dagger\hat{\sigma}'_-)}$ provides the Hamiltonian in the new frame,
\begin{equation}\label{eq:frequency_shift}
    \hat{H}'_\mathrm{dress}=\left(\frac{\Omega_R}{2}+\frac{g^2}{2\Delta_R}\right)\hat{\sigma}'_z+\left(\omega_h+\frac{g^2}{\Delta_R}\hat{\sigma}'_z\right)\hat{a}^\dagger \hat{a}.
\end{equation}
In this regime, the qubit and the harmonic oscillator do not exchange energy, but the resonance frequency of the harmonic oscillator gets shifted by $\pm\frac{g^2}{\Delta_R}$ depending on the state of the qubit. This shift of the resonance frequency can be used as a readout mechanism to measure the state of the qubit~\cite{Lyyra:2024,Vanner:2011,Vanner:2013,Muhonen:2019,Blais:2004,Blais:2007}.

\section{Steady state of the dressed qubit}\label{sec:stationary_state}
At $T=0$, a bare qubit relaxes to its ground state due to environment-induced damping. However, this is not the case for a dressed qubit, where the qubit-heat bath interaction no longer describes energy quanta exchange with the heat bath. In this section, we derive the steady state of a dressed qubit coupled to a harmonic oscillator. Some derivations were performed using \textit{Mathematica}, the corresponding notebook is available at~\cite{programs}. To simplify the calculations, we adopt the rotating wave approximation taken in Appendix~\ref{sec:spring effect}. At low temperatures, the harmonic oscillator is thermalized in its ground state. The qubit-harmonic oscillator interaction [see Eq.~\eqref{eq:RWA Hamiltonian}] couples the states $|1,n=0\rangle$ and $|0,n=1\rangle$, where the second index denotes the oscillator's quantum number. Consequently, we can restrict our analysis to the subspace spanned by $\{|1,n=0\rangle, |0,n=0\rangle, |0,n=1\rangle\}$. In this basis, the Hamiltonian takes the form
\begin{equation}
    \hat{H}_\mathrm{dress}=
    \begin{pmatrix}
        \frac{\Omega_R}{2} & 0 & g\\
        0 & -\frac{\Omega_R}{2} & 0\\
        g & 0 & -\frac{\Omega_R}{2}+\omega_h
    \end{pmatrix}\; .
\end{equation}

The steady state may be computed from the quantum master equation~\eqref{eq:QME} by setting $\dot{\hat{\rho}}=0$. Transforming the Lindblad operators into the dressed-qubit basis and restricting to the relevant subspace, the equation reduces to a linear system for the density matrix elements, allowing for a closed-form solution.

We first examine the steady state of a dressed qubit in the absence of coupling to the harmonic oscillator $g=0$. In this case, the $|0,n=1\rangle$ state is decoupled from the states with $n=0$, so the steady state will only depend on $|1,n=0\rangle$ and $|0,n=0\rangle$. For weak damping $\gamma_q\ll\Omega_R$, taking the $\gamma_q\rightarrow 0^+$ limit, the density matrix describing the steady state simplifies to
\begin{equation}\label{eq:stationary_decoupled}
    \hat{\rho}_\mathrm{st}=\frac{1}{2}(|0\rangle\langle0|+|1\rangle\langle1|)\; ,
\end{equation}
where we have omitted the harmonic oscillator's quantum number for simplicity. Equation~\eqref{eq:stationary_decoupled} describes a qubit in a maximally mixed state. The qubit dressing symmetrizes the $|0\rangle$ and $|1\rangle$ states (and, more generally, all pure states), causing them to decay into an equidistant maximally mixed state.

Next, we analyze a dressed qubit coupled to a harmonic oscillator. In the dispersive regime $g\ll|\Delta_R|$. Moreover, we consider weak damping rates satisfying $\gamma_q,\gamma_h \ll g$. Expanding the closed-form solution to zeroth order in the ratio of the damping rates to $g$ and subsequently to second order in $g$, the elements of the density matrix become
\begin{subequations}\label{eq:stationary_coupled}
\begin{align}
    \rho_{1,1}&=\frac{1}{2}-\frac{\gamma_q+4\gamma_h}{\tilde{\gamma}}\frac{g^2}{\Delta_R^2}\\
    \rho_{2,2}&=\frac{1}{2}+\frac{4\gamma_h-\gamma_q}{\tilde{\gamma}}\frac{g^2}{\Delta_R^2}\\
    \rho_{3,3}&=\frac{2\gamma_q}{\tilde{\gamma}}\frac{g^2}{\Delta_R^2}\\
    \rho_{1,3}&=\frac{g}{2\Delta_R},\; \rho_{1,2}=\rho_{2,3}=0\; .
\end{align}
\end{subequations}
where $\tilde{\gamma}=\frac{16\gamma_q\gamma_h}{5\gamma_q+4\gamma_h}$. As shown in Eq.~\eqref{eq:stationary_coupled}, the coupling between the $|1,n=0\rangle$ and $|0,n=1\rangle$ states introduces a correction of order $O(g^2/\Delta_R^2)$ to their population in the density matrix, decreasing $\rho_{1,1}$ and increasing $\rho_{3,3}$. These corrections reduce the qubit polarization $\langle\hat{\sigma}'_z\rangle$ ($\langle\hat{\sigma}_x\rangle$ in the bare qubit basis used in the main text)
\begin{equation}\label{eq:stationary_state}
    \langle\hat{\sigma}'_z\rangle=-\frac{2(\gamma_q+4\gamma_h)}{\tilde{\gamma}}\frac{g^2}{\Delta_R^2}
\end{equation}
In summary, the steady state of a dressed qubit is a maximally mixed state. However, coupling to the harmonic oscillator drives the system toward a steady state that is closer to the ground state of the dressed qubit.

\section{Quantum bus between two dressed qubits}\label{sec:iSWAP}
In this section, we consider two driven qubits coupled to a harmonic oscillator. The harmonic oscillator works as a quantum bus between qubits, entangling both qubits~\cite{Blais:2004,Liu:2006,Srinivasa:2024}. The Hamiltonian for the system can readily be obtained by generalizing Eq.~\eqref{eq:RWA Hamiltonian},
\begin{equation}\label{eq:RWA Hamiltonian 2}
    \hat{H}_\mathrm{dress}=\sum_j\frac{\Omega_{R,j}}{2}\hat{\sigma}_z^{\prime j}+\omega_h\hat{a}^\dagger \hat{a}+\sum_j g_j(\hat{\sigma}_-^{\prime j}\hat{a}^\dagger+\hat{\sigma}_+^{\prime j}\hat{a})\; ,
\end{equation}
where $j=1,2$ is the qubit index. Following the same steps as in Appendix~\ref{sec:spring effect}, applying a Schrieffer-Wolff transformation of the form $\hat{U}=e^{\sum_j g_j/\Delta_{R,j}(\hat{a}\hat{\sigma}_+^{\prime j} - \hat{a}^\dagger\hat{\sigma}_-^{\prime j})}$, in the dispersive regime the Hamiltonian acquires an effective qubit-qubit interaction term:
\begin{equation}\label{H_eff}
\begin{aligned}
    \hat{H}'_\mathrm{dress}=\sum_j\left(\frac{\Omega_{R,j}}{2}+\frac{g_j^2}{2\Delta_{R,j}}\right)&\hat{\sigma}_z^{\prime j} +\left(\omega_h+\sum_j\frac{g_j^2}{\Delta_{R,j}}\hat{\sigma}_z^{\prime j}\right)\hat{a}^\dagger \hat{a}\\
    &+\sum_j\frac{g_j g_{\bar{j}}}{2\Delta_{R,j}}(\hat{\sigma}_+^{\prime j}\hat{\sigma}_-^{\prime\bar{j}}+\hat{\sigma}_-^{\prime j}\hat{\sigma}_+^{\prime\bar{j}})\; ,
\end{aligned}
\end{equation}
where $\bar{j}$ is the qubit opposite to $j$.

If the transition frequencies and the coupling constants of both qubits are the same, the interaction term (second line of Eq.~\eqref{H_eff}) commutes with the noninteracting Hamiltonian part $\hat{H}_0$, so the time evolution operator is given by~\cite{Zheng:2000,Sorensen:1999,Raimond:2001}
\begin{equation}
    \hat{U}(t)=e^{-i \hat{H}_0 t}
    \begin{pmatrix}
        1 & 0 & 0 & 0\\
        0 & \cos{\frac{g^2 t}{\Delta_R}} & -i\sin{\frac{g^2 t}{\Delta_R}} & 0\\
        0 & -i\sin{\frac{g^2 t}{\Delta_R}} & \cos{\frac{g^2 t}{\Delta_R}} & 0\\
        0 & 0 & 0 & 1
    \end{pmatrix}\; .
\end{equation}
If the qubit-qubit interaction time is fixed to $t=\frac{\pi|\Delta_R|}{4g^2}$ and the harmonic oscillator remains in the ground state, $\hat{U}(t)$ corresponds to a $\sqrt{\mp i\mathrm{SWAP}}$-gate depending on the sign of the detuning $\Delta_R$ [see Eq.~\eqref{eq:sqrt(iSWAP)}], up to one-qubit gates. The interaction time can be controlled by tuning the drive amplitudes $\Omega_{R,j}(t)$ of the qubits, such that an appropriate pulse length tunes the dressed qubits in resonance with the harmonic oscillator. The $\sqrt{i\mathrm{SWAP}}$-gate is equivalent to a CNOT gate up to one-qubit gates, so the interaction in Hamiltonian~\eqref{H_eff} is sufficient for universal quantum computation~\cite{Barenco:1995}.

\begin{figure}[h!]
    \centering
    \includegraphics[width=0.95\columnwidth]{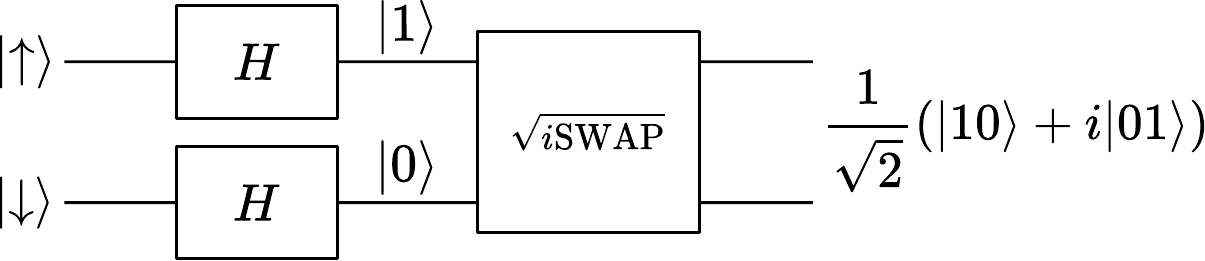}
    \caption{Generation of a Bell state in the computational basis.}
    \label{fig:Bell_generation}
\end{figure}

\section{Creating Bell states with dressed qubits}\label{sec:Bell_states}
Bell states are maximally entangled quantum states of two qubits. They violate the Bell inequality, disproving the local hidden-variable theories. Most quantum algorithms rely on entanglement, so Bell states are a key resource in quantum computing. In this section, we show how to generate Bell states in dressed qubits using a $\sqrt{i\mathrm{SWAP}}$-gate, Eq.~\eqref{eq:sqrt(iSWAP)}. The steady state of a dressed qubit is not a suitable initial state for quantum computing because it is not a pure state. Instead, we may apply a Zeeman field to initialize the qubit in its bare ground state $|\!\downarrow\rangle$ via thermal relaxation. The computational basis states $\{|0\rangle,|1\rangle\}$ ($\hat{\sigma}_x$ eigenstates) can easily be obtained by applying a Hadamard gate. Applying the $\sqrt{i\mathrm{SWAP}}$-gate to two qubits with opposite parity will generate a Bell state, as shown in Fig.~\ref{fig:Bell_generation}.

\bibliography{biblio}

\end{document}